\documentclass[11pt,a4paper]{article}
\usepackage{bm}
\usepackage{amsmath}

\usepackage{jheppub}

\newcommand{\be}{\begin{equation}}
\newcommand{\ee}{\end{equation}}
\newcommand{\bea}{\begin{eqnarray}}
\newcommand{\eea}{\end{eqnarray}}

\newcommand{\calF}{\mathcal{F}}
\newcommand{\calL}{\mathcal{L}}
\newcommand{\calM}{\mathcal{M}}

\title{Black hole thermodynamics, conformal couplings, and $R^2$ terms}
\author[a]{Mariano Chernicoff,}
\author[b,f]{Mario Galante,}
\author[b,c,d]{Gaston Giribet,}
\author[b]{Andres Goya,}
\author[b]{Matias Leoni,}
\author[e]{Julio Oliva,}
\author[b]{and Guillem Perez-Nadal}
\affiliation[a]{Departamento de F\'{\i}sica, Facultad de Ciencias,
Universidad Nacional Aut\'onoma de M\'exico; A.P. 70-542, M\'exico D.F. 04510, M\'exico}
\affiliation[b]{Departamento de F\'{\i}sica, Universidad de Buenos Aires and IFIBA - CONICET; {Ciudad Universitaria, pabell\'on 1 (1428) Buenos Aires.}}
\affiliation[c]{Universit\'e Libre de Bruxelles and International Solvay Institutes; {Campus Plaine C.P. 231 B-1050, Bruxelles, Belgium.}}
\affiliation[d]{Instituto de F\'isica, Pontificia Universidad Cat\'olica de Valpara\'{\i}so; {Casilla 4950, Valpara\'{\i}so, Chile.}}
\affiliation[e]{Departamento de F\'{i}sica, Universidad de Concepci\'on; Casilla 160-C, Concepci\'on, Chile.}
\affiliation[f]{Van Swidenderen Institute for Particle Physics and Gravity, University of Groningen; {Nijenborgh 4, 9747 AG Groningen, The Netherlands.}}

\date{\today}

\abstract{Lovelock theory provides a tractable model of higher-curvature gravity in which several questions can be studied analytically. This is the reason why, in the last years, this theory has become the favorite arena to study the effects of higher-curvature terms in the context of AdS/CFT correspondence. Lovelock theory also admits extensions that permit to accommodate matter coupled to gravity in a non-minimal way. In this setup, problems such as the backreaction of matter on the black hole geometry can also be solved exactly. In this paper, we study the thermodynamics of black holes in theories of gravity of this type, which include both higher-curvature terms, $U(1)$ gauge fields, and conformal couplings with matter fields in $D$ dimensions. These charged black hole solutions exhibit a backreacting scalar field configuration that is regular everywhere outside and on the horizon, and may exist both in asymptotically flat and asymptotically Anti-de Sitter (AdS) spaces. We work out explicitly the boundary action for this theory, which renders the variational problem well-posed and suffices to regularize the Euclidean action in AdS. We also discuss several interrelated properties of the theory, such as its duality symmetry under field redefinition and how it acts on black holes and gravitational wave solutions.} 

\emailAdd{mchernicoff@ciencias.unam.mx}
\emailAdd{mario@df.uba.ar}
\emailAdd{gaston@df.uba.ar}
\emailAdd{af.goya@df.uba.ar}
\emailAdd{leoni@df.uba.ar} 
\emailAdd{julioolivazapata@gmail.com}
\emailAdd{guillem@df.uba.ar}

\begin{document}

\maketitle
\flushbottom
%\setlength{\parskip}{8pt}
%\newpage

%%%%%%%%%%%%%%%%%

\section{Introduction}

Being the natural extension of general relativity (GR) to higher dimensions \cite{Lovelock}, Lovelock theory provides a tractable model of higher-curvature gravity in which several problems can be solved exactly \cite{BoulwareDeser}. This is the reason why, in the last years, this theory has been used as a model to investigate the effects of higher-curvature corrections in the context of holography \cite{Brigante}-\cite{Edelstein} and also to study the consistency of higher-curvature corrections \cite{Maldacena, GiribetEdelstein, GiribetEdelstein2, Reall}. Lovelock theory admits an extension that permits to accommodate matter conformally coupled to gravity \cite{Oliva1}, which also results in a model in which several problems can be solved analytically. For instance, in such setup one can explicitly work out the details of black holes with backreacting matter both in asymptotically flat and in asymptotically $\rm (Anti-)$de Sitter spacetimes and in arbitrary number of dimensions \cite{Giribet1}. This has been used in \cite{Giribet2, Galante1} to study Hawking-Page type phase transitions of black holes decorated with scalar fields in 5-dimensional Anti-de Sitter (AdS) space, proposing a setup in which such problem, of relevance for holographic applications to the description of critical phenomena, can be solved explicitly. The extended phase space thermodynamics of these asymptotically AdS, charged black holes with conformal scalars, shows a rich structure presenting Van der Waals behaviour as well as reentrant phase transitions \cite{Hennigar}. In this paper, we will extend the study of the thermodynamics of these {\it hairy} black holes in several directions: we will consider the black hole solutions of the theory that, apart from the coupling with conformal matter, also include Lovelock type quadratic-curvature corrections to the pure gravity action; we will also consider the coupling of $U(1)$ gauge fields. For such black holes, we will perform a thermodynamical analysis in arbitrary dimension $D>4$ and discuss the qualitative difference between the case $D>5$ and the case $D=5$ studied in \cite{Giribet2, Galante1}. We will consider the case of cosmological constant of arbitrary sign, including also solutions that are asymptotically flat. The study of thermodynamics in the latter case demands a careful consideration of the boundary terms of the theory, which we construct here explicitly. We work out both the boundary action needed for the variational principle to be correctly defined and the local counter-terms that suffice to regularize the Euclidean action in AdS. We show the consistency of our analysis by computing the thermodynamical quantities in different ways; namely, by computing quantities such as the free energy both by background subtraction and by the introduction of counter-terms to regularize the on-shell action. Our analysis provides a set of examples in which analytic hairy black hole solutions non-minimally coupled to matter and including higher-curvature terms can be solved exactly.  

The paper is organized as follows: in section 2, we review Lovelock theory of gravity and its generalization that includes conformal couplings with scalar matter, proposed in \cite{Oliva1}. In section 3, we discuss the black hole solutions to this theory, which exhibit backreaction of the scalar field and charges under additional gauge fields. In section 4, we work out the boundary action and show the consistency of such terms. In section 5, we use the boundary action to compute the conserved charges and study the thermodynamics of the black holes. This serves to confirm and generalize previous results in the literature. In section 6, we discuss the duality symmetry that the theory exhibits under frame changing, discussing this in relation to the black hole solutions and to other solutions such as $pp$-waves.

\section{Higher-curvature terms and conformal couplings}
\label{theory}

\subsection{Lovelock theory}

Lovelock theory is defined by the action
\be\label{aL}
I_{\{a\}}[g]=\int_{\mathcal M} d^Dx \sqrt{-g}\sum^{[\frac{D-1}{2}]}_{k=0}\frac{1}{2^k}
\delta^{\mu_1\nu_1\dots \mu_k\nu_k}_{\alpha_1\beta_1\dots \alpha_k\beta_k }
\ a_k \ R^{\ \ \ \ \alpha_1\beta_1}_{\mu_1\nu_1}\dots R^{\ \ \ \ \alpha_k\beta_k}_{\mu_k\nu_k}
  \ + I_{B,\{a\}} \,,
\ee
where the generalized Kronecker tensor is defined as
\be
\delta^{\mu_1\nu_1\dots \mu_k\nu_k}_{\alpha_1\beta_1\dots \alpha_k\beta_k } \equiv (2k)! \ \delta^{\mu_1}_{[\alpha_1}
\delta^{\nu_1}_{\beta_1} \ldots\ \delta^{\mu_k}_{\alpha_k} \delta^{\nu_k}_{\beta_k ]} \,,
\ee
$R^{\ \mu}_{\nu\ \lambda\delta}$ are the components of the Riemann curvature tensor, and the symbol $[n]$ stands for the integer part of $n$. The term $I_{B,\{a\}}$ in (\ref{aL}) is a boundary contribution, defined on the boundary $\partial {\mathcal M}$ of the $D$-dimensional manifold ${\mathcal M}$. This term, which will be an important ingredient in our discussion, will be written in section 3. Let us first focus on the theory in the bulk.

The theory defined by (\ref{aL}) is the most general theory of pure gravity with vanishing torsion whose fields equations are given by a rank-2 symmetric tensor, covariantly conserved and of second order in the metric $g$ \cite{Lovelock}. Consequently, in $D\leq 4$ this theory coincides with Einstein theory \cite{Lanczos, Lovelock2}. In contrast, in dimension $D>4$ this results in a generalization of general relativity (GR) in which the cosmological Einstein-Hilbert action is augmented with higher-curvature terms of order ${\mathcal O}(R^k)$ with $k\leq [(D-1)/2]$. Explicitly, up to cubic terms in the curvature, action (\ref{aL}) reads
\begin{multline}
I_{\{a\}}[g] = \int_{\mathcal M}d^D x\sqrt{-g}  \Big{[}a _{0}+a _{1}R+a _{2}\Big{(}
R^{2}+R_{\alpha \beta \mu \nu }R^{\alpha \beta \mu \nu }-4R_{\mu \nu }R^{\mu
\nu }\Big{)}\\ + a_{3} \Big{(}R^{3}+ 3RR^{\mu \nu \alpha \beta }R_{\alpha \beta \mu \nu
}- 12RR^{\mu \nu }R_{\mu \nu }+ 24R^{\mu \nu \alpha \beta }R_{\alpha \mu
}R_{\beta \nu } + 16R^{\mu \nu }R_{\nu \alpha }R_{\mu }^{\alpha } \\+ 24R^{\mu \nu \alpha \beta }R_{\alpha \beta \nu \rho }R_{\mu }^{\rho
} + 8R_{ \alpha \rho }^{\mu \nu }R_{\nu \sigma }^{\alpha \beta }R_{\mu \beta }^{\rho \sigma }+2R_{\alpha \beta \rho \sigma }R^{\mu \nu \alpha \beta }R_{ \mu \nu }^{\rho \sigma } \Big{)} + \ldots \Big{]}
\end{multline}
where the ellipses stand for higher order and boundary terms.

The field equations derived from varying this action with respect to the metric $g$ read 
\begin{align}
G_{\mu}^{\nu}   \equiv \sum_{k=0}^{\left[  \frac{D-1}{2}\right]  }\frac{a_{k}%
}{2^{k+1}}\delta_{\mu\beta_{1} \ldots \beta_{2k}}^{\nu\alpha_{1} \ldots \alpha_{2k}%
}R_{\ \ \ \ \alpha_{1}\alpha_{2}}^{\beta_{1}\beta_{2}} \ldots R_{\ \ \ \ \ \ \ \alpha
_{2k-1}\alpha_{2k}}^{\beta_{2k-1}\beta_{2k}}\ =0 \,, \label{elG}
\end{align}
which, indeed, are of second order in $g$, although not of degree one in the second derivatives $\partial^2 g$.

Action (\ref{aL}) is also interesting from the mathematical point of view because of its connection with topology. In fact, each term in the Lovelock Lagrangian corresponds to the dimensional extension of the Chern-Weil topological invariants, in the same way as the Einstein-Hilbert action can be thought of as the dimensional extension of the 2-dimensional Euler characteristic. This observation provides one with a systematic way of constructing the boundary terms $I_{B,\{a\}}$.

The collection of real numbers $a_0, a_1, \ldots  a_k$ in (\ref{aL}), to which the subindex $\{a\}$ on the right hand side of (\ref{aL}) refers, are coupling constants of the theory that, while in principle independent, are expected to correspond to a unique length scale when the theory is thought of as a truncation of a fundamental one. Actually, the quadratic Lovelock action appears in next-to-leading contributions to the low energy effective action of string theory \cite{Zwiebach} and, in that context, the different orders ${\mathcal O}(R^k)$ appear as $(\alpha ' )^{k-1}$ corrections. Both bosonic and heterotic strings exhibit such quadratic term, in a particular frame, and it also appears in Calabi-Yau compactifications of M-theory to 5 dimensions. In string and M-theory, however, the higher-curvature terms appear in convolution with the dilaton or moduli fields, and this is better described by the coupling of Lovelock theory to matter. 

\subsection{Conformal couplings}

Among the different ways to couple matter to Lovelock theory, there is one that results particularly interesting for providing a tractable model and exhibiting conformal symmetry: the generalization of Lovelock theory proposed in \cite{Oliva1} consists of spin-0 matter coupled to gravity through a conformaly invariant coupling between a real scalar field $\phi $ and the dimensionally extended Euler densities $\sim {\mathcal O}(R^k)$ of arbitrary order $k\leq [(D-1)/2]$ in $D$ spacetime dimensions. This yields a theory whose field equations are of second order and presents quite interesting properties such as self-duality under frame changing. This can be regarded as a natural generalization of Lovelock gravity with couplings to matter, and in this sense it is also a generalization of Horndeski \cite{Horndeski} and Galileon \cite{Galileon} theories. 

The theory admits non-minimal couplings that can be conveniently expressed in terms of a four-rank tensor $S^{\ \mu}_{\nu \ \ \lambda\delta}$ defined by
\be\label{Stensor}
S_{\mu\nu} ^{\ \ \lambda\delta}\equiv \phi^2R^{\ \ \lambda\delta}_{\mu\nu}+\frac{4}{s}\phi\delta^{[\lambda}_{[\mu}\nabla_{\nu]}\nabla^{\delta]}\phi +\frac{4(1-s)}{s^2}\delta^{[\lambda}_{[\mu}\nabla_{\nu]}\phi\nabla^{\delta]}\phi -\frac{2}{s^2}\delta^{[\lambda}_{[\mu}\delta^{\delta]}_{\nu]}\nabla_{\rho}\phi\nabla^{\rho}\phi \,,
\ee
where $s$ is a real parameter different from zero. This tensor can be shown to transform homogeneously under the Weyl transformation
\be\label{conformaltrans}
g_{\mu \nu }\rightarrow e^{2\Omega}g_{\mu \nu }, \qquad \phi \rightarrow e^{s\Omega}\phi \,.
\ee
More precisely, under Weyl rescaling (\ref{conformaltrans}) the tensor (\ref{Stensor}) transforms as follows 
\be
S_{\mu\nu} ^{\ \ \lambda\delta} \rightarrow e^{2(s-1)\Omega }S_{\mu\nu} ^{\ \ \lambda\delta} \,.
\ee

In terms of tensor (\ref{Stensor}) the general action of Lovelock theory coupled to conformal matter takes the form \cite{Oliva1}
\begin{align}\label{theoryaction}
I[g,\phi ]=\int_{\mathcal M} d^Dx \sqrt{-g}\sum^{[\frac{D-1}{2}]}_{k=0}\frac{1}{2^k}
\delta^{\mu_1\nu_1\dots \mu_k\nu_k}_{\alpha_1\beta_1\dots \alpha_k\beta_k }
\Big{(} & a_k \ R^{\ \ \alpha_1\beta_1}_{\mu_1\nu_1}\dots R^{\ \ \alpha_k\beta_k}_{\mu_k\nu_k}
+\nonumber\\
& b_k\phi^{m_k}  S^{\ \ \alpha_1\beta_1}_{\mu_1\nu_1}\dots S^{\ \ \alpha_k\beta_k}_{\mu_k\nu_k}  \Big{)} + I_B \,,
\end{align}
where $a_k$ and $b_k$ represent coupling constants that are in principle arbitrary, and $m_k=-(D-2k)/s-2k$, $s$ being the conformal weight of the scalar field.

Choosing $s=-(D-2)/2$ and expanding the action (\ref{theoryaction}), one finds the more familiar form of a conformally coupled field theory
\be\label{theoryactionmatter}
I[g,\phi ]= \int_{\mathcal M} d^Dx \sqrt{-g}
\Big{(}  \frac{1}{16\pi G} R - \frac{\Lambda}{8\pi G} - \partial_{\mu }\phi \partial^{\mu} \phi - \frac{(D-2)}{4(D-1)} \phi^{2}R - \frac{\lambda}{D!} \phi^{\frac{2D}{(D-2)}} + \, \ldots \Big{)} \,,
\ee
where the ellipses stand for terms of order ${\mathcal O}(R^2)$ and higher, and where we have renamed the coupling constants conveniently, 
\be
a_0=-\frac{\Lambda}{8\pi G} \,,\ \ \ \ a_1=\frac{1}{16\pi G}\,,\ \ \ \ b_0=-\frac{\lambda}{D!}\,,\ \ \ \ b_1=-1 \,.
\ee

The field equations coming from varying (\ref{theoryaction}) with respect to the metric can be written in the form
\begin{equation}\label{E210}
G_{\mu\nu}=T_{\mu\nu}  \,,
\end{equation}
with $G_{\mu\nu}$ given by (\ref{elG}) and $T_{\mu\nu}$ given by
\begin{align}
T_{\mu}^{\nu}   =\sum_{k=0}^{\left[  \frac{D-1}{2}\right]  }\frac{b_{k}%
}{2^{k+1}}\phi^{m_k}\delta_{\mu\beta_{1} \ldots \beta_{2k}}^{\nu\alpha
_{1} \ldots \alpha_{2k}}S_{\ \ \ \ \alpha_{1}\alpha_{2}}^{\beta_{1}\beta_{2}%
} \ldots S_{\ \ \ \ \ \ \ \ \alpha_{2k-1}\alpha_{2k}}^{\beta_{2k-1}\beta_{2k}} \,.
\end{align}

Despite the presence of higher-curvature terms in the action, these equations are of second order in the metric. The same happens with the equation for the scalar field $\phi $, which takes the form
\begin{equation}\label{onyel}
\sum_{k=0}^{[\frac{D-1}{2}]} \frac{ \left(2k-D\right) b_{k} }{ 2^{k}s } \phi^{m_{k}-1} \delta^{\mu_1\nu_1\dots \mu_k\nu_k}_{ \alpha_1\beta_1\dots \alpha_k\beta_k} S^{\ \ \ \ \alpha_1\beta_1}_{\mu_1\nu_1} \dots S^{\ \ \ \ \alpha_k\beta_k}_{\mu_k\nu_k}=0 \,.
\end{equation}

It is possible to show, using (\ref{onyel}), that the trace of  $T_{\mu\nu}$ vanishes on-shell, which is consistent with the conformal invariance of the matter action. It can also be shown that the matter part of the action, meaning the terms in (\ref{theoryaction}) with couplings $b_k$, can be thought of as a theory of pure gravity for the rescaled metric $\tilde{g}_{\mu \nu }\equiv \phi ^{-2/s} g_{\mu \nu}$. This follows from the fact that the Riemann tensor associated to the metric $\tilde{g}_{\mu\nu}$ reads
\begin{equation}\label{RS}
	\tilde{R}^{\ \ \ \mu\nu}_{\alpha \beta }= \phi^{2/s-2}{S}^{\ \ \ \mu\nu}_{\alpha \beta } \, ,
\end{equation}
therefore, the matter part of action (\ref{theoryaction}) can be written as a pure gravity action
\be\label{E214}
I_\text{matt}[g,\phi ]=\int_{\mathcal M} d^Dx \sqrt{-\tilde{g}}\sum^{[\frac{D-1}{2}]}_{k=0}\frac{1}{2^k}
\delta^{\mu_1\nu_1\dots \mu_k\nu_k}_{\alpha_1\beta_1\dots \alpha_k\beta_k }
\ b_k \ \tilde{R}^{\ \ \ \ \alpha_1\beta_1}_{\mu_1\nu_1}\dots \tilde{R}^{\ \ \ \ \alpha_k\beta_k}_{\mu_k\nu_k}
 \ +I_{B,\text{matt}} \,,
\ee
and viceversa. Here, $I_{B,\text{matt}}$ refers to the piece of the boundary action that depends on the matter content; see the next section. In other words,
\be
I_\text{matt}[g,\phi ]= I_{\{b\}}[\phi ^{-2/s} g] \,.
\ee
This remark will be relevant later when we will study the properties of the solutions dual to the $D$-dimensional black holes, in section 6. The self-duality under frame exchange permits to associate to a given solution $(g_{\mu \nu }, \phi )$ of the field equations a {\it dual} solution $(\tilde{g}_{\mu \nu }, \phi^{-1})$ which, in principle, exhibits qualitatively different geometrical properties. For instance, consider the theory with $b_0 = a_0 =0$, for which a solution is given by
\be
ds^2= -dt^2+dr^2+r^2d\Sigma_{{D-2} }^2 \,,  \ \ \ \ \phi = Q \ r + \phi_0 \,,
\ee
with $d\Sigma^2_{D-2 }$ being the metric on the unit ($D-2$)-sphere, and with $Q$ and $\phi_0$ constants. That is, this is flat space $g_{\mu \nu }=\eta_{\mu \nu}$ with a linear scalar field $\phi \sim r$. The metric of the dual solution, $\tilde{g}$, is actually conformally flat, but certainly not flat, while $\tilde{\phi }$ results singular at $r=-\phi_0/Q$. There are other interesting solutions, such as stealth solutions \cite{Oliva1}, and self-dual $pp$-waves which, under the frame changing, turn out to be diffeomorphic equivalent to themselves. The symmetry under interchanging $g_{\mu \nu } \leftrightarrow \tilde{g}_{\mu \nu }$ requires, of course, a specific relation between the couplings $a_i$ and $b_i$. Such selfdual points of the parameter space, the solutions exhibit special features. For instance, in $D=3$ dimensions the black hole solution \cite{MTZ} exists only at such selfdual point. The higher-dimensional black holes of \cite{Giribet1}, in contrast, do not require the selfdual coupling condition to exist. We will discuss this symmetry in detail in section 6.

\section{Boundary terms and variational principle}

\subsection{Non-minimal coupling to matter and $R^k$ terms}

Let us now construct the boundary terms $I_{B}$. These terms are required for the variational principle to be well defined: the contribution $I_B$ suffices to make the action (\ref{theoryaction}) functionally differentiable. The terms corresponding to the Lovelock Lagrangian, the purely gravitational part of the action (\ref{theoryaction}), can be derived from the dimensional extension of the boundary terms of characteristic classes. Their form is known. Interestingly, it turns out that the form of the piece $I_{B, \text{matt}}$, corresponding to the conformal coupled matter, can also be inferred in a similar way based on the observation that, as discussed in section 2, the full action (\ref{theoryaction}) can be rewritten as the sum of two Lovelock actions, one for the metric $g$ and one for the rescaled metric $\tilde{g}$; see (\ref{E214}). In other words,
\begin{equation}\label{action2}
I[g,\phi ]=I_{\{a\}}[g]+I_{\{b\}}[\tilde{g}] \,,
\end{equation}
where $\tilde{g}_{\mu\nu}=\phi^{-2/s}g_{\mu\nu}$. Recall that $I_{\{c\}}$ stands for the action of Lovelock gravity in $D$ dimensions (\ref{aL}) with coupling constants $c_0, c_1, \dots \ c_{[(D-1)/2]}$; namely
\begin{equation}
I_{\{c\}}[g]=\sum_{k=0}^{[\frac{D-1}{2}]}\frac{c_k}{2^k}\int_\calM d^Dx\sqrt{-g}\,\delta^{\mu_1\nu_1\dots \mu_k\nu_k}_{\alpha_1\beta_1\dots \alpha_k\beta_k}\,R^{\ \ \ \ \alpha_1\beta_1}_{\mu_1\nu_1}\dots R^{\ \ \ \ \alpha_k\beta_k}_{\mu_k\nu_k} + I_{B,\{c \}} \,.
\label{lov}
\end{equation}

This also tells us that the boundary term that renders the full action functionally differentiable is given by
\begin{equation}\label{boundary}
I_B=I_{B,\{a\}}[g]+I_{B,\{b\}}[\phi^{-2/s}g] \,,
\end{equation}
where $I_{B, \{c\}}$ is the boundary term for the Lovelock action with coupling constants $c_0$, $... $, $c_{[(D-1)/2]}$; namely 
\begin{alignat}{2}
I_{B,\{c\}}[g_{\mu\nu}]=\sum_{k=1}^{[\frac{D-1}{2}]}c_k\sum_{l=0}^{k-1}\zeta_{kl}\int_{\partial \calM}d^{D-1}x & \sqrt{|h|}\,\delta^{\sigma\mu_1\nu_1\dots \mu_{k-1}\nu_{k-1}}_{\rho\alpha_1\beta_1\dots \alpha_{k-1}\beta_{k-1}}\times\nonumber\\
\times &K^{\rho}_\sigma K^{\alpha_1}_{\mu_1}\dots K^{\beta_l}_{\nu_l}
\hat R^{\ \ \ \ \ \ \ \alpha_{l+1}\beta_{l+1}}_{\mu_{l+1}\nu_{l+1}}\dots \hat R^{\ \ \ \ \ \ \ \alpha_{k-1}\beta_{k-1}}_{\mu_{k-1}\nu_{k-1}} \,,
\end{alignat}
where $h$ is the metric induced by $g$ on the boundary $\partial\calM$, $K$ is the extrinsic curvature of the boundary according to $g$, $\hat R$ is the Riemann tensor made out of $h$ and
\begin{equation}
\zeta_{kl}=\frac{(-1)^l\,2k!}{2^{k-1-l}(2l+1)l!(k-1-l)!} \, ;
\end{equation}
see for instance \cite{Miskovic1} and reference therein. The extrinsic curvature of the boundary according to the metric $\phi^{-2/s}g$ is
\begin{equation} \label{KL}
\bar K^{\mu}_{\nu}=\phi^{1/s-1}L^{\mu}_{\nu} \,,
\end{equation}
with
\begin{equation}
L^{\mu}_{\nu}=\phi\, K^{\mu}_\nu-\frac{1}{s}\,\Pi^{\mu}_\nu\, n^\alpha\nabla_\alpha\phi \,.
\end{equation}
Here, $\Pi$ denotes the orthogonal projection onto the boundary and $n$ is the unit vector normal to the boundary, pointing outward if the boundary is timelike and inward if it is spacelike. Using equations (\ref{RS}) and (\ref{KL}), we can write down a more explicit expression of the boundary term (\ref{boundary}); namely
\begin{alignat}{2}\label{IB}
I_B=\sum_{k=1}^{[\frac{D-1}{2}]}\sum_{l=0}^{k-1}\zeta_{kl}\int_{\partial \calM}d^{D-1}x&\sqrt{|h|}\,\delta^{\sigma\mu_1\nu_1\dots \mu_{k-1}\nu_{k-1}}_{\rho\alpha_1\beta_1\dots \alpha_{k-1}\beta_{k-1}}\times\nonumber\\
\times&\left(a_kK^{\rho}_\sigma K^{\alpha_1}_{\mu_1}\dots K^{\beta_l}_{\nu_l}
\hat R^{\ \ \ \ \ \ \alpha_{l+1}\beta_{l+1}}_{\mu_{l+1}\nu_{l+1}}\dots \hat R^{\ \ \ \ \ \ \alpha_{k-1}\beta_{k-1}}_{\mu_{k-1}\nu_{k-1}}+\right.\nonumber\\
&\left.+b_k\phi^{m_k+1}L^{\rho}_\sigma L^{\alpha_1}_{\mu_1}\dots L^{\beta_l}_{\nu_l}\hat S^{\ \ \ \ \alpha_{l+1}\beta_{l+1}}_{\mu_{l+1}\nu_{l+1}}\dots \hat S^{\ \ \ \ \alpha_{k-1}\beta_{k-1}}_{\mu_{k-1}\nu_{k-1}}\right) \,,
\end{alignat}
where $\hat S$ is the tensor obtained via Eq.~(\ref{Stensor}) from the metric and the scalar field induced on $\partial {\mathcal M}$. Notice from Eqs.~(\ref{theoryaction}) and (\ref{IB}), together with Eqs.~(\ref{Stensor}) and (\ref{KL}), that the complete action of the theory is analytic in the scalar field if and only if $m_k$ is a positive integer for all values of $k$ whose associated parameters $b_k$ are non-vanishing. This is achieved, for example, when $-1/s$ is a positive integer not smaller than $2N/(D-2N)$ with $N = [(D-1)/2]$. In that case, there is a unique analytic extension of the action to non-positive scalar fields, which is given also by Eqs.~(\ref{theoryaction}) and (\ref{IB}).

\subsection{Boundary terms and $R^2$ terms}

In the next sections we will be concerned with the $k=1$ and $k=2$ terms of the boundary action \eqref{IB}. Their explicit expressions are
\begin{alignat}{2}
I_{B}^{(k=1)}&=2\int_{\partial\calM}d^{D-1}x\sqrt{|h|}\left(a_1 K+b_1\phi^{m_1+1} L\right) \,, \\
I_{B}^{(k=2)}&=4\int_{\partial\calM}d^{D-1}x\sqrt{|h|}\left[a_2\left(J-2\hat G^{\mu\nu}K_{\mu\nu}\right)+b_2\phi^{m_2+1}\left(J_L-2\hat G_{\hat S}^{\mu\nu}L_{\mu\nu}\right)\right] \,,
\end{alignat}
respectively, where $J$ is the trace of the tensor
\begin{equation}\label{E312}
J_{\mu\nu} = \dfrac{1}{3} \left( 2KK_{\mu\alpha}K^{\alpha}_{\ \nu} + K_{\alpha\beta}K^{\alpha\beta}K_{\mu\nu} - 2K_{\mu\alpha}K^{\alpha\beta}K_{\beta\nu} - K^2K_{\mu\nu} \right) \,,
\end{equation}
$\hat G$ is the Einstein tensor of the metric induced on the boundary, $J_L$ is obtained from $J$ by replacing $K$ by $L$, and $\hat G_{\hat S}$ is obtained from $\hat G$ by replacing $\hat R$ by $\hat S$. The matter part of these terms is given more explicitly by
\begin{equation}\label{BdyS1}
I_{B,\text{matt}}^{(k=1)} = 2b_1 \int_{\partial \calM} d^{D-1}x \, \sqrt{|h|} \phi^{-(D-2)/s} \left( K - \dfrac{D-1}{s \, \phi} \calL_{n}\phi \right) \,,
\end{equation}
and
%where $\calL_n\phi = n^{\alpha} \nabla_{\alpha} \phi$.
%For $k=2$ we have
\begin{align}\label{BdyS2}
I_{B,{\text{matt}}}^{(k=2)} &=4b_2 \int_{\partial \calM} d^{D-1}x \, \sqrt{|h|} \, \phi^{-(D-4)/s} \Big( J - 2 \hat{G}_{\mu\nu}K^{\mu\nu} + \hspace*{0.5cm} \nonumber\\
	& (D-3) \, \Bigg\lbrace \, (D-2) \, \Bigg[ \dfrac{D-1}{3s^3} \left( \dfrac{\calL_n \phi}{\phi} \right)^3 - \dfrac{1}{s^2} \left(\dfrac{\calL_n \phi}{\phi} \right)^2 K - \nonumber\\
	&  \dfrac{1}{s^2} \left( \dfrac{2 \hat{\nabla}^2\phi}{\phi} - \dfrac{D-1-2(1-s)}{s} \dfrac{\hat{\nabla}\phi\cdot\hat{\nabla}\phi}{\phi^2} \right)  \dfrac{\calL_n \phi}{\phi} - \nonumber\\
	& \dfrac{1}{s^2} \dfrac{\hat{\nabla}\phi \cdot \hat{\nabla}\phi}{\phi^2} \, K\,  \Big] - \dfrac{1}{s}\dfrac{\calL_n \phi}{\phi} \left( \hat{R} + K_{\mu\nu}K^{\mu\nu} - K^2 \right)- \nonumber\\
	&  \dfrac{2}{s}\left( K^{\mu\nu} - Kh^{\mu\nu} \right) \left( \dfrac{\hat{\nabla}_{\mu}\hat{\nabla}_{\nu}\phi}{\phi} + \dfrac{(1-s)}{s} \dfrac{\hat{\nabla}_{\mu}\phi\hat{\nabla}_{\nu}\phi}{\phi^2}  \right) \,  \Bigg\rbrace \, \Bigg) \,,
\end{align}
where $\calL_n\phi = n^{\alpha} \nabla_{\alpha} \phi$ and $\hat{\nabla}$ is the covariant derivative on the boundary associated to $h$.

%where $\hat{\nabla}$ are the covariant derivatives projected onto the boundary, $\hat{G}_{\mu\nu}$ is the Einstein tensor of the induced metric $h_{\mu\nu}$, and $J$ is the trace of the tensor $J_{\mu\nu}$, which is given by
%\begin{equation}
%	J_{\mu\nu} = \dfrac{1}{3} \left( 2KK_{\mu\alpha}K^{\alpha}_{\ \nu} + K_{\alpha\beta}K^{\alpha\beta}K_{\mu\nu} - 2K_{\mu\alpha}K^{\alpha\beta}K_{\beta\nu} - K^2K_{\mu\nu} \right) \,.
%\end{equation}

In section 5, we will resort to the expression of the boundary terms (\ref{BdyS1})-(\ref{BdyS2}) to compute the conserved charges and thermodynamical quantities of black holes with backreacting scalar field. Let us first introduce, in the next section, the black hole geometries and discuss their main properties.

\section{The black hole solution and backreaction}
\label{bhsolution}

\subsection{Black holes with $R^k$ terms and matter}

In contrast to what happened in 3 and 4 dimensions, where asymptotically (Anti)-de Sitter black hole solutions in Einstein gravity conformally coupled to scalar field matter have been known explicitly for long time \cite{MTZ, Martinez1, Henneaux, Martinez2}, in higher dimensions no-go results for the existence of such similar static spherically solutions had been reported until recently \cite{Martinez:nogo}. It has been proven in \cite{Giribet1}, however, that such no-go results could be circumvented by adding the conformal couplings of the type discussed in section 2. We have already seen that these new conformal couplings involve higher-curvature terms together with higher-derivatives of the scalar field, all arranged in a way that the field equations turn out to be conformally invariant as well as of second order \cite{Oliva1}. 

In \cite{Giribet1}, explicit black hole solutions were found in these theories in arbitrary dimension $D>4$, for horizons of either positive or negative curvature, and for arbitrary sign of the cosmological constant, generalizing in this way the results of \cite{Martinez1, Martinez2}. There are special relations between the coupling constants $b_0, b_1, ...\ b_{[(D-1)/2]}$ for which the static spherically symmetric solution to the field equations (\ref{E210})-(\ref{onyel}) takes the notable simple form
\be\label{solutionhigher}
ds^2=-F(r)dt^2+F^{-1}(r)\ dr^2+r^2d\Sigma^2_{D-2}  \,,\quad  \text{and} \qquad \phi(r)=\frac{N}{r} \,.
\ee
We will consider $r\in \mathbb{R}_{\geq 0}$, $t\in \mathbb{R}$, and $d\Sigma^2_{D-2}$ being the metric of a $(D-2)$-dimensional Euclidean space of constant curvature. More precisely, we will consider this space to be the unit $(D-2)$-sphere of volume $Vol_{\Sigma}=2\pi^{(D-1)/2}/\Gamma ((D-1)/2)$. The metric function $F(r)$ is root of the polynomial equation \cite{Wheeler}
\be\label{polynomialeq}
\sum^{[\frac{D-1}{2}]}_{k=0}a_k\frac{(D-1)!}{(D-2k-1)!}\Big{(}\frac{1-F(r)}{r^2}\Big{)}^k=\frac{M(D-1)}{Vol_{\Sigma}r^{D-1}}+\frac{Q_0}{r^D} \,,
\ee
and $N$ satisfies the following constraints
\be\label{const1}
\sum^{[\frac{D-1}{2}]}_{k=1} k\tilde{b}_k N^{2-2k}=0 \,.
\ee
For (\ref{solutionhigher})-(\ref{polynomialeq}) to be a solution, the coupling constants have to satisfy
\be\label{const2}
\sum^{[\frac{D-1}{2}]}_{k=0} (D(D-1)+4k^2)\tilde{b}_k N^{-2k}=0 \,,
\ee
with $\tilde{b}_k=b_k(D-1)!/(D-2k-1)!$. Notice that (\ref{const1})-(\ref{const2}) admit particular solutions with $b_{k>2}=0$ for generic $D\geq 5$. In (\ref{polynomialeq}), $M$ is an arbitrary integration constant corresponding to the black hole mass, while $Q_0$ is fixed in terms of the coupling constants by
\be\label{charge}
Q_0=\sum^{[\frac{D-1}{2}]}_{k=0} (D-2k-1)\tilde{b}_k N^{D-2k} \,.
\ee

We see from (\ref{polynomialeq}) that the metrics of these black holes receive a contribution from the conformally coupled scalar field. The scalar field configuration, on the other hand, remains finite everywhere outside and on the horizon. The particular dependence on $Q_0$ in the metric can be understood as the change of gravitational energy contribution 
\begin{equation}\label{jhgflkhf}
\frac{(D-1)}{Vol_{\Sigma } r^{D-3}}\ M \to \frac{(D-3)}{Vol_{\Sigma } r^{D-1}} \left( M+ \frac{Vol_{\Sigma }}{(D-1)}\frac{Q_0}{r} \right) \,,
\end{equation} 
which corresponds to the shifting $M\to M+\Delta M $ with $\Delta M= \int d\Sigma \int_{0^+}^r dr \ r^{D-2}T_{0}^{\ 0}(r) $ being the contribution of the field $\phi $ to the energy (the divergence in the limit $r\to 0^+$ in the radial integral is absorbed by renormalizing $M$). Since the scalar matter is coupled to gravity in a conformal invariant manner, the lack of energy scale demands the component of the energy-momentum tensor $T_{0}^{\ 0}(r)$ to be proportional to $ Q_0/r^{D}$, and this precisely results in (\ref{jhgflkhf}).

\subsection{Charged $R^2$ black holes}

Equation (\ref{polynomialeq}) generically has $(D-1)/2$ real roots. Only one of such solutions is physically sensible, leading to the spherically symmetric solution of GR in the limit in which $a_{n>1}$ tends to zero. Here, we are interested in studying the behaviour of the solution including terms ${\mathcal O}(R^2)$ in the purely gravitational part of the action, and therefore we set $a_{n>2}=0$. In such case, equation (\ref{polynomialeq}) takes the form
\be
a_0+a_1\frac{(D-1)!}{(D-3)!}\Big{[}\frac{1-F(r)}{r^2}\Big{]}+a_2\frac{(D-1)!}{(D-5)!}\Big{[}\frac{1-F(r)}{r^2}\Big{]}^2-\frac{M(D-1)}{Vol_{\Sigma}r^{D-1}}-\frac{Q_0}{r^D}=0 \,, \label{E46}
\ee
and solving for the metric function $F(r)$ explicitly, one obtains
\be
F(r) = 1+\frac{r^2}{4\tilde{\alpha}}-\frac{r^2}{4\tilde{\alpha}} H(r) \,,
\ee
with
\begin{equation}\label{HH}
\begin{split}
H^2(r) &= 1-\frac{8\tilde{\alpha}a_0}{(D-1)(D-2)a_1}+\frac{8\tilde{\alpha}M}{Vol_{\Sigma}(D-2)a_1r^{D-1}}+\frac{8\tilde{\alpha}Q_0}{(D-1)(D-2)a_1r^D}
 \, ;
\end{split}
\end{equation}
where, for convenience, we have defined 
\be
\tilde{\alpha} =\frac{{a_2(D-4)(D-3)}}{{2a_1}} \,.
\ee

The inclusion of a $U(1)$ gauge field in the theory (\ref{theoryaction}), minimally coupled to the curvature, produces an additional contribution to the right hand side of (\ref{HH}), namely
\be\label{cucurucho}
H^2(r) \to H^2(r) - \frac{8\tilde{\alpha } Q_1^2}{(D-2)(D-3)a_1 Vol_{\Sigma }r^{2D-4}}  \,,
\ee
where $Q_1$ is the electric charge of the black hole; that is, the gauge field behaves as $A(r)\propto Q_1r^{3-D}dt$. Topological black hole solutions also exist \cite{cai1, cai2}.

Equation (\ref{HH}) is quadratic, permitting both signs for $H(r)$. This accounts for the two roots of $F(r)$. Positive $H(r)$ in the limit $\tilde{\alpha }\to 0$ yields
\begin{multline}
F(r)\simeq 1-\frac{16\pi G M}{Vol_{\Sigma } (D-2) r^{D-3}}+\frac{16\pi G Q_1^2}{Vol_{\Sigma } (D-2)(D-3) r^{2D-6}}
\\-\frac{16\pi G Q_0}{(D-1)(D-2)r^{D-2}}- \frac{2\Lambda r^2}{(D-1)(D-2)} + {\mathcal O}(\tilde{\alpha }) \,,
\end{multline}
where we have defined
\be
a_0 \equiv -\frac{\Lambda }{8\pi G} \,.
\ee
This leads to the GR solution, with the second term on the right hand side being the Newtonian potential of the Schwarzschild-Tangherlini solution, the third and four being the Reissner-Nordstr\"{o}m and the CFT matter contributions, respectively (notice that both coincide in the case $D=4$). The fifth term corresponds to the cosmological constant. Negative $H(r)$, in contrast, in the limit $\tilde{\alpha }\to 0$ yields
\begin{multline}
F(r)\simeq 1-\frac{16\pi G (-M)}{Vol_{\Sigma } (D-2) r^{D-3}}-\frac{16\pi G Q_1^2}{Vol_{\Sigma } (D-2)(D-3) r^{2D-6}}\\+\frac{16\pi G Q_0}{(D-1)(D-2)r^{D-2}}- 
\frac{2\Lambda_{\text{eff}} \ r^2}{(D-1)(D-2)} + {\mathcal O}(\tilde{\alpha }) \,,
\end{multline}
with the effective cosmological constant
\be
\Lambda_{\text{eff}} \equiv -\Lambda -\frac{(D-1)(D-2)}{4\tilde{\alpha }} \,,
\ee
which tends to infinity when $\tilde{\alpha}\to 0$, provided $\Lambda $ remains fixed. This is the reason why the solution $H(r)>0$ is regarded as the only physical sensible solution. In \cite{BoulwareDeser} the divergence of $\Lambda_{\text{eff}}$ in the limit $\tilde{\alpha}\to 0$ is rephrased as the theory having {\it its own cosmological constant problem}. We will consider the case of $H(r)$ being positive with $\tilde{\alpha }>0$ in order to avoid other pathological features (see (\ref{entropy}) below). When $\tilde{Q}_0=0$, solution (\ref{solutionhigher}) reduces to the well-known Boulware-Deser solution \cite{BoulwareDeser}. 

There is another case that deserves special attention. This is
\be
a_1 = \frac{8\tilde{\alpha } a_0}{(D-1)(D-2)} \,, \label{mmmm}
\ee
which corresponds to the case in which the two roots of (\ref{E46}) coincide. In $D=5$ the theory satisfying (\ref{mmmm}) corresponds to the so-called Chern-Simons gravity \cite{Zanelli}, a point in the parameter space in which the theory exhibits a symmetry enhancement and enjoys gauge invariance. This means that condition (\ref{mmmm}) should not be regarded as a fine tuning. 

To express the general solution, it will also be convenient to define the variables
\be
\qquad \tilde{\Lambda}=-\frac{a_0}{(D-1)(D-2)a_1} \,, \quad \tilde{M}=\frac{M}{Vol_{\Sigma}(D-2)a_1} \,, \quad \tilde{Q}_0=\frac{Q_0}{(D-1)(D-2)a_1} \,.
\ee
In general, the solution above presents a horizon at $r_+$, with
\be
-\tilde{\Lambda}r^{2D-4}_{+} + r^{2D-6}_{+} +2\tilde{\alpha}r^{2D-8}_{+} - \tilde{M}r_+^{D-3} -\tilde{Q}_0r_+^{D-4}+\tilde{Q}_1^2=0 \,,
\ee
where 
\be
\tilde{Q}_1^2=-\frac{Q_1^2}{(D-2)(D-3)a_1Vol_{\Sigma }} \,.
\ee
From this, we can write the mass in terms of $r_+$; namely
\be
M=Vol_{\Sigma }(D-2)a_1\left( -\tilde{\Lambda}r^{D-1}_{+} + r^{D-3}_{+} +2\tilde{\alpha}r^{D-5}_{+}  +\tilde{Q}_1^2r_+^{3-D} - {\tilde{Q}_0}{r_+}^{-1} \right) \label{HJK} \,.
\ee

Notice that, if $Q_1=0$, the last term in (\ref{HJK}), if positive, renders the mass spectrum of small black holes unbounded from below. This pathology suggests to consider only the black holes with $\tilde{Q}_0\leq 0$, and excluding the solutions with $\tilde{Q}_0>0$ for being unphysical. It is worth mentioning that, even if additional matter contributions are coupled to the theory in a way that (\ref{HJK}) receives a UV correction that dominates over $-\tilde{Q}_0/r_+$ at short distance (for instance, by turning on the $U(1)$ charge (\ref{cucurucho}) with $Q_1\neq 0$), the solution with $\tilde{Q}_0>0$ still presents other pathologies, such as negative entropy configurations \cite{Galante1}, and therefore in what follows we will restrict ourselves to the case $\tilde{Q}_0<0$. However, it is still interesting to take a look at the special case $M=0$ with $\tilde{Q}_0>0$. Such a solution is also a black hole. For $\tilde{\alpha }=\tilde{\Lambda }=0$, it yields
\be
F(r)=1-\frac{4\tilde{Q}_0}{r^{D-2}} \,,
\ee
which, provided $\tilde{Q}_0 >0$, presents a horizon at $r_+ = (4\tilde{Q}_0)^{1/(D-2)}$.

Another aspect that deserves attention in order to see whether a given solution makes sense is the following: while the positive roots of $F(r)$ give the location of the horizons of the solution (with the biggest value $r_+$ being the location of the external event horizon), the roots of $H(r)$ define the surfaces on which the metric may become non-real. Therefore, in order for the metric to be real, we must require $H(r)$ not to have roots greater than $r_+$. 

There is another case that is special, which corresponds to the extremal configurations; namely, the one satisfying
\be
\tilde{Q}_0r^{3-D}_+ + (D-3)r_+ + 2\tilde{\alpha}(D-5)r^{-1}_+ - \tilde{\Lambda}(D-1)r^3_+ -(D-3)\tilde{Q}_1^2 r_+^{7-2D} =0 \,.
\ee
This solution corresponds to double roots for $F(r_{\pm })=0$, and in the near horizon limit adopts the form AdS$_2\times S^{D-2}$.

\section{Black hole thermodynamics}
\label{thermo}

\subsection{Thermodynamics of $R^2$ black holes}

The black hole solutions discussed above have non-trivial thermodynamical properties. For the AdS black holes in $D=5$, this has been studied in \cite{Giribet2, Galante1}. Motivated by the applications that such tractable model could have from the holographic perspective, in \cite{Giribet2} the Hawking-Page type phase transition of such 5-dimensional (hairy) black holes in AdS space were investigated, arriving to the conclusion that a transition yielding a non-vanishing scalar field charge takes place only at high temperature. The introduction of a $U(1)$ gauge field was introduced in \cite{Galante1}, altering the short distance behaviour in a way that a new type of instabilities emerges at low temperatures.       

As we will see here, higher-dimensional black holes have some properties qualitatively different from the $D=5$ case. The Hawking temperature of black hole solutions (\ref{solutionhigher}) with $a_{n>2}=0$ is given by 
\begin{align}\label{temperature}
T(r_+) = \frac{1}{4\pi(4\tilde{\alpha}+r^2_+)} \Big( \tilde{Q}_0 r^{3-D}_+ + (D-3)r_+ 
+ & 2\tilde{\alpha}(D-5)r^{-1}_+ - \nonumber\\ 
& \tilde{\Lambda}(D-1)r^3_+ -(D-3)\tilde{Q}_1^2 r_+^{7-2D}  \Big) \,.
\end{align}
In the following, we will consider the case $\tilde{Q}_1=0$, since the presence of the $U(1)$ charge, at large $r_+$, does not introduce any new qualitative feature with respect to the Wiltshire electrically charged black holes in quadratic Lovelock theory \cite{Wiltshire}.

Temperature (\ref{temperature}) presents different features depending on the parameters and it is necessary to consider different cases. Let us begin discussing the different regimes of (\ref{temperature}): at large $r_+$, the temperature correctly reproduces the behaviour of black holes in GR; namely $T(r_+)\simeq -(D-1)\tilde{\Lambda }r_+/(4\pi ) + {\mathcal O}(r_+^{-1}) $. The difference appears for small values of $r_+$. At short distances and in the case $\tilde{\alpha }\neq 0$, (\ref{temperature}) behaves like
\be\label{temperature2}
T(r_+)\simeq \frac{1}{16\pi \tilde{\alpha}} \left[ \tilde{Q}_0r^{3-D}_+ + (D-3) r_+ + 2\tilde{\alpha}(D-5)r^{-1}_+  \right] \,.
\ee

Therefore, analyzing the limit $r_+\to 0$ of (\ref{temperature2}) demands first to distinguish between three different cases, depending on which of the three terms in (\ref{temperature2}) is the relevant one in each case: first, consider the case $\tilde{Q}_0\neq 0$, for which, independently of the dimension $D$, the temperature diverges as
\be\label{temperature3}
T(r_+)\simeq \frac{1}{16\pi \tilde{\alpha}} \frac{\tilde{Q}_0}{r^{D-3}_+} + {\mathcal O}(1/r_+) \,. 
\ee

This is qualitatively different from the case $\tilde{Q}_0=0$ (provided $\tilde{Q}_1=0$, as the effects of the electric charge do dominate over (\ref{temperature3}) at short distances). When $\tilde{Q}_0$ vanishes, we have to distinguish between the case $D>5$ and the special case $D=5$. For $\tilde{\alpha }\neq 0$, $\tilde{Q}_0=0$ and $D> 5$, the short distance behavior is $T(r_+)\simeq (D-5)/(8\pi \tilde{\alpha} {r_+}) + {\mathcal O}(r_+)$; while for $\tilde{\alpha }\neq 0$, $\tilde{Q}_0=0$ and $D=5$, the short distance behavior is $ T(r_+)\simeq (D-3)r_+/(16\pi \tilde{\alpha}) + {\mathcal O}(r_+^3)$. This means that these black holes in 5 dimensions present positive specific heat at short scales, what eventually yield remnants. In general, Lovelock black holes with ${\mathcal O}(R^k)$ terms exhibit this feature in $D=2k+1$ dimensions.

The temperature of $\tilde{\alpha} =0$ black holes, on the other hand, presents qualitatively different features at short distances with respect to the behaviours described above. While for $\tilde{Q}_0=0$ one merely obtains the general relativity result $ T(r_+)\simeq 1/(2\pi  {r_+}) + {\mathcal O}(r_+)$. In the case $\tilde{Q}_0\neq 0$, in contrast, the short distance behaviour changes drastically; it reads
\be\label{temperature7}
T(r_+)\simeq \frac{\tilde{Q}_0}{4\pi  {r}^4_+} + {\mathcal O}(1/r_+) \,.
\ee

Notice that here we did not yet discuss the sign of $Q_0$, so the short distance behaviors above may represent qualitatively diverse cases. For example, we can achieve finite size black holes with zero temperature, and black holes in AdS with arbitrarily low temperature. All  this depends on the sign and absolute value of $\tilde{Q}_0$ in relation to the other parameters. Furthermore, many of the configurations may exhibit pathologies. We also mentioned that solutions with positive values of $\tilde{Q}_0$ may have mass $M$ not bounded from below and/or negative entropy. The latter is due to the fact that the entropy can acquire a constant additive contribution ($S_0$ in (\ref{entropy}) below) whose sign is sensitive to the sign of $Q_0$. The menu of possible behaviors is even more diverse if $\tilde{\alpha }\neq 0$, since the function $H(r)$ may in general present, for instance, roots larger than $r_+$, as we already mentioned. We will avoid discussing pathological cases such as branch singularities of this type.

From the integration of the first law of black hole thermodynamics, we obtain from (\ref{temperature}) the entropy 
\be\label{entropy}
S =  4\pi \, Vol_{\Sigma}(D-2) \, a_1 \, \left[ \frac{r^{D-2}_+}{D-2}+\frac{4\tilde{\alpha}r^{D-4}_+}{D-4} \right] +S_0 \,.
\ee

This result can be alternatively obtained by means of the Wald entropy formula \cite{Wald}. $S_0$ is the constant contribution we mentioned above, which appears due to the presence of higher-curvature couplings. Such contribution, which is not zero due to the coupling of conformal matter to higher-curvature terms, will be written explicitly below for the case $D=5$. Notice that, while the first term in (\ref{entropy}) corresponds to the Bekenstein-Hawking area law, the entropy formula also have other contributions; namely
\begin{equation}
S=\frac{A}{4G}+{\mathcal O}(\tilde{\alpha } r^{D-4}) \,.
\end{equation}

In order to avoid negative values of (\ref{entropy}) for positive $r_+$, we restrict our analysis to the cases $\tilde{\alpha }\geq 0$ and $\tilde{Q}_0\leq 0$.

In the following sections, we will explicitly compute the conserved charges and other thermodynamical quantities in terms of the Euclidean action.

\subsection{The 5-dimensional black holes}

Let us now discuss in detail the particular case of 5-dimensional black holes. For convenience, here we prefer to work with a different conformal weight for the scalar field: we rescale $\phi\to \phi^{1/5}$, so that now $\phi$ has conformal weight $s=-1/5$. For this value of $s$, the limit $\phi\to 0$ is completely safe. Consistently, we rescale $N$ in such a way that the scalar field configuration is $\phi(r)=N/r^{1/5}$. Let us also define $\ell \equiv 1/\sqrt{|\tilde\Lambda |}$, $a_0 = -\Lambda / 8 \pi G$ and $a_1 = 1 / 16\pi G$, and consider the case $\tilde{\alpha }=0$. From the definition of $\tilde\Lambda$ above, $\ell$ is seen to be the radius of the AdS$_5$ space.

\begin{align}\label{ID5}
I[g,\phi ]  = \int_{\mathcal M} d^{5}x\sqrt{-g} \Big{[} \frac{1}{16\pi G}R-\frac{\Lambda}{8\pi G} + & b_{0}\phi^{25}+b_{1}\phi^{13} S+ \nonumber\\
&	b_{2}\phi^{}\left( S_{\mu\nu\alpha\beta}S^{\mu\nu\alpha\beta
}-4S_{\mu\nu}S^{\mu\nu}+S^{2} \right)  \Big{]} \,,
\end{align}
where $S_{\mu\nu}=S_{\ \mu\rho\nu}^{\rho}$ and $S=S_{\mu}^{\ \mu}$.

Black hole solution with backreacting scalar field in $D=5$ is given by
\begin{equation}\label{5Dsolf}
	F(r) = 1 - \dfrac{\tilde{M}}{r^2} - \dfrac{\tilde{Q}_0}{r^3} + \dfrac{r^2}{\ell^2} \, ,
\end{equation}
with the coupling constants $b_i$, the scalar field charge $\tilde{Q}_0$ and scalar field strength $N$ satisfying the following relations
\begin{equation}\label{5Dsolb}
	b_2 = \dfrac{9}{10} \dfrac{b_1^2}{b_0} \,, \quad
	\tilde{Q}_0 = \dfrac{64\pi G}{6} \epsilon b_1 \left(-\dfrac{18}{5}\dfrac{b_1}{b_0}\right)^{3/2} \,, \quad
	N = \epsilon \left(-\dfrac{18}{5}\dfrac{b_1}{b_0}\right)^{1/10} \, ,
\end{equation}
where $\epsilon =-1,0,+1$. It is instructive to compare the condition on the couplings (\ref{5Dsolb}), namely
\begin{equation}
b_0b_2=\frac{9}{10}b^2_{1} \,,
\end{equation}
with the causality constraints found in the study of AdS/CFT in $D=5$ Lovelock theory \cite{Brigante}, which reads
\begin{equation}
a_0a_2<\frac{27}{50}a^2_{1} \,.
\end{equation} 
That is, the quadratic coupling to matter, when seen as a Gauss-Bonnet action for the dual metric $\tilde{g}$, is outside the causality bound.

The scalar field profile is, as said,
\begin{equation}\label{scalarfieldprofile}
    \phi(r) = \dfrac{N}{r^{1/5}} \,,
\end{equation}
which is regular everywhere, except at the origin. It backreacts on the metric via the term proportional to $1/r^3$, which does not spoil its AdS$_5$ asymptotic at large $r$. In fact, at large $r$ one finds in particular
\be
g_{tt} \simeq - \frac{r^2}{\ell ^2} + {\mathcal O}(1) \,, \ \ \ \ \ \ \ g_{rr} \simeq  \frac{\ell ^2}{r^2} + {\mathcal O}(r^{-4}) \, .
\ee
The associated Hawking temperature is given by
\begin{equation}\label{5Dsoltemp}
	T =  \dfrac{r_+}{\ell^2\pi } + \dfrac{1}{2\pi \,r_+} + \dfrac{\tilde{Q}_0}{4\pi \,r_+^4} \,.
\end{equation}

%\section{Boundary terms and conserved charges}

Now, we will make explicit use of the boundary terms \eqref{IB}. More specifically, we will evaluate the Euclidean action of the theory in the black hole solution. In order to do so, it is necessary the explicit expression of \eqref{IB} for $D=5$. The appropriate boundary terms for the action \eqref{ID5}, including the adequate coefficients, are given by the expressions \eqref{BdyS1} and \eqref{BdyS2} for $D=5$ and $s=-1/5$,
\begin{equation}\label{BdyS1unquinto}
I^{(k=1)}_{B,\text{matt}}[g,\phi] = 2b_1 \int_{\partial {\mathcal M}} d^{4}x \, \sqrt{|h|} \phi^{15} \left( K + \dfrac{20}{\phi} \calL_{n}\phi \right) \,,
\end{equation}
and
\begin{align}\label{BdyS2unquinto}
	I^{(k=2)}_{B,\text{matt}} [g,\phi] = 4b_2\int_{\partial {\mathcal M}} d^{4}x \, \sqrt{|h|} \, \phi^{5} & \Big( J - 2 \hat{G}_{\mu\nu}K^{\mu\nu} + 10 \, \Big\lbrace \, 15 \, \Big[ -\dfrac{20}{3} \Big( \dfrac{\calL_n \phi}{\phi} \Big)^3 + \Big(\dfrac{\calL_n \phi}{\phi} \Big)^2 K + \nonumber\\
	&    \Big( \dfrac{2 \hat{\nabla}^2\phi}{\phi} + 200 \dfrac{\hat{\nabla}\phi\cdot\hat{\nabla}\phi}{\phi^2} \Big)  \dfrac{\calL_n \phi}{\phi} + \dfrac{\hat{\nabla}\phi \cdot \hat{\nabla}\phi}{\phi^2} \, K\,  \Big] + \nonumber\\
	& 2\Big( K^{\mu\nu} - Kh^{\mu\nu} \Big) \Big( \dfrac{\hat{\nabla}_{\mu}\hat{\nabla}_{\nu}\phi}{\phi} - 6 \dfrac{\hat{\nabla}_{\mu}\phi\hat{\nabla}_{\nu}\phi}{\phi^2}  \Big)  \,  + \nonumber\\
	&   \, \Big( \hat{R} + K_{\mu\nu}K^{\mu\nu} - K^2 \Big) \dfrac{\calL_n \phi}{\phi}\Big\rbrace \Big{)} \,,
\end{align}
where $\hat{\nabla}$ is the covariant derivative projected onto the boundary.

Below, we will study the thermodynamics of both asymptotically AdS and asymptotically flat black holes. Let us begin with the former.

\subsection{Asymptotically AdS black holes}

We want to compute the on-shell action for the black hole configuration \eqref{5Dsolf}-\eqref{5Dsolb} in an ensemble at fixed temperature and fixed volume. Therefore, the adequate thermodynamical potential to analyze is the Helmholtz free energy $\calF$, which is obtained from the Euclidean action $I_E = \beta \calF$. Considering the boundary terms defined above, the free energy of AdS$_5$ black holes computed by substracting the Euclidean action of thermal AdS$_5$ and taking care of the appropriate red-shift factor when matching the Euclidean time, turns out to be given by
\begin{equation}\label{FAdS}
	\calF_{\rm (AdS)} = -\frac{1}{8}\,\frac{\pi \,r_+^4}{G\ell^2} + \frac{1}{8}\,\frac{\pi \,r_+^2}{G} + \frac{5}{4}\, \frac{\pi \,\tilde{Q}_0 r_+}{G\ell^2} + \frac{1}{8}\,\frac{\pi \,\tilde{Q}_0}{G r_+} + \frac{5}{16}\,\frac{\pi \,\tilde{Q}_0^2}{G r_+^4} \,,
\end{equation}
which exactly agrees with the result previously obtained in \cite{Giribet2}. This result for the free energy manifestly shows the existence of Hawking-Page type transition for different values of $\tilde{Q}_0$. The entropy, computed from \eqref{FAdS}, is
\begin{eqnarray}
\label{SAdS}
S_{\rm (AdS)} =  \beta^2 \dfrac{d\calF_{\rm (AdS)}}{d\beta} = \frac12\,\dfrac{\pi^2 r_+^3}{G} - \dfrac{5}{4}\,\dfrac{\tilde{Q}_0 \pi^2}{G} \,,
\end{eqnarray}
while the mass is given by
\begin{eqnarray}
\label{MAdS}
M_{\rm (AdS)} =  \calF_{\rm (AdS)} + \beta \dfrac{d\calF_{\rm (AdS)}}{d\beta} = \dfrac{3}{8}\, \dfrac{\pi \, \left( r_+^5 + r_+^3 \ell^2 - \tilde{Q}_0\ell^2 \right)}{G\ell^2 r_+} \, .
\end{eqnarray}

These quantities can be shown to satisfy the first law $dM_{\rm (AdS)} = T\ dS_{\rm (AdS)}$ and to match the results of the previous sections, yielding the GR values for $\tilde{Q}_0=0$; namely
\begin{equation}\label{M.Sch.AdS}
	S_{\rm (AdS)}^{(\tilde{Q}_0=0)} = \frac12\,\dfrac{\pi^2 r_+^3}{G} \,, \ \ \ \ \ 
	M_{\rm (AdS)}^{(\tilde{Q}_0=0)} = \frac{3}{8}\,\dfrac{\pi \,r_+^2 \left( r_+^2 + \ell^2 \right) }{G \ell^2 } \,.
\end{equation}

As a crosscheck, and as a second application of the boundary terms \eqref{IB}, let us compute the on-shell action but now using counterterms to regularize the action directly, instead of resorting to the background subtraction. To do so, we should supplement the action \eqref{ID5}-\eqref{BdyS2unquinto} with the appropriate counterterms to obtain a finite result. The relevant counterterms are given by
\begin{equation}\label{CounterTerms}
	I_0^{\rm (ct)} = k_0 \int_{\partial {\mathcal M}} d^4x \sqrt{h} \,, \quad I_1^{\rm (ct)} = k_1 \int_{\partial {\mathcal M}} d^4x \sqrt{h} \hat{R} \, ,
\end{equation}
with the appropriate values of the coefficients $k_0$ and $k_1$ to remove the IR divergences. These values are
\begin{equation}\label{CounterTermsCoeffs}
	k_0 = \dfrac{3}{8} \dfrac{1}{\ell \pi G} \,, \ \ \quad k_1 = \dfrac{1}{32} \dfrac{\ell}{\pi G} \,.
\end{equation}
The free energy computed in this manner matches (\ref{FAdS}) except for an additive constant. This is consistent, since such constant exactly agrees with the free energy of thermal AdS renormalized by the counterterms (\ref{CounterTerms}). Let us address the case of asymptotically flat black holes.

\subsection{Asymptotically flat black holes}

In the case of asymptotically flat spacetimes, the Euclidean action is found to be
\begin{equation}\label{FFlat}
I_{E} = \beta \left( \dfrac{1}{8}\,\dfrac{\pi \,r_+^2}{G} + \dfrac{1}{8}\,\dfrac{\pi \,\tilde{Q}_0}{G r_+} + \dfrac{5}{16}\,\dfrac{\pi \, \tilde{Q}_0^2}{G r_+^4}\right) \, ,
\end{equation}
from which one computes the entropy
\begin{eqnarray}
\label{SFlat}
S_{\rm (Flat)} = \frac12\,\dfrac{\pi^2 r_+^3}{G} - \dfrac{5}{4}\,\dfrac {\tilde{Q}_0 \pi^2}{G} \,,
\end{eqnarray}
and the mass 
\begin{eqnarray}
\label{MFlat}
M_{\rm (Flat)} = \dfrac{3}{8}\,\dfrac{\pi \,r_+^2}{G} - \dfrac{3}{8}\,\dfrac{\pi \,\tilde{Q}_0}{G r_+} \, .
\end{eqnarray}

These expressions also satisfy the first law, and in the case $\tilde{Q}_0=0$ yield the known results of GR. Therefore, both in the case of asymptotically AdS and asymptotically flat black holes, we find that the action including the boundary terms constructed in \eqref{IB} gives the correct contribution. This completes the definition of the theory proposed in \cite{Oliva1}.

\section{Frame duality, gravitational waves and black holes}

\subsection{Frame duality}

As mentioned in section 2, given the relation between the Riemann tensor and the
conformally covariant tensor $S_{\lambda\rho}^{\ \ \mu\nu}$, there is a frame changing transformation that maps solutions of the theory (\ref{E210}) defined with a specific set of coupling constants $\{a_0, a_1, ...$ $a_{[(D-1)/2]};$ $b_0, b_1, ...$ $b_{[(D-1)/2]}\}$ into solutions of the theory defined by different values of the couplings. For Lagrangians linear in the curvature, this type of duality transformation was explored in \cite{Shapiro:1995kt}. In this
section we will discuss the details of this duality for the higher-curvature theory. The complete bulk action is (\ref{theoryaction}) and can be schematically written as follows
\begin{equation}
I_{\text{bulk}}\left[  g,\phi\right]  =\int_{\mathcal{M}}d^{D}x\sqrt{-g}%
\sum_{k=0}^{\left[  \frac{D-1}{2}\right]  }\frac{1}{2^{k}}\left(  a_{k}%
\delta^{\left(  k\right)  }R^{\left(  k\right)  }\left(  g\right)  +b_{k}%
\phi^{m_{k}}\delta^{\left(  k\right)  }S^{\left(  k\right)  }\left(
\phi,g\right)  \right)  \,,\label{shortaction}%
\end{equation}
where $\delta^{(k)}R^{(k)}$ represent the contraction of the generalized Kronecker tensor and the product of $k$ Riemann tensors appearing in (\ref{aL}); respectively for $\delta^{(k)}S^{(k)}$. 

Performing the map $\left(  g_{\mu\nu},\phi\right)  =\left(  (\xi / \tilde{\phi} )^{2/s}
 \tilde{g}_{\mu\nu},\phi_{0}(\xi / \tilde{\phi})  \right)  $ (we have introduced the constants $\xi$ and
$\phi_{0}$ to properly keep track of dimensions), the Riemann and the conformally covariant tensors result in
\begin{align}
R_{\mu\nu}^{\ \ \lambda\rho}     & =\tilde{\phi}^{2/s-2}\xi^{-2/s} \ \tilde{S}_{\mu\nu
}^{\ \ \lambda\rho}  \,, \nonumber\\
S_{\mu\nu}^{\ \ \lambda\rho}   & = {\tilde{\phi}}^{2/s-2}{\xi}^{2-2/s}  \phi_{0}^{2} \ \tilde{R}_{\mu\nu}^{\ \ \lambda\rho
} \,,
\end{align}
where $\tilde{S}$ stands for the tensor constructed with the scalar $\tilde{\phi }$ and the metric $\tilde{g}$, and $\tilde{R}$ is the Riemann tensor associated to $\tilde{g}$. This means that action (\ref{shortaction}) acquires the following form
\begin{align}\label{shortactiondual}
& I_{\text{bulk}}\left[  (\xi / \tilde{\phi})^{2/s}  \tilde{g},\phi_{0}(\xi / \tilde{\phi})  \right] =  \nonumber\\
& \int_{\mathcal{M}}d^{D}x\sqrt{-\tilde{g}}\sum_{k=0}^{\left[  \frac{D-1}%
{2}\right]  }\frac{1}{2^{k}}\left(  b_{k}\phi_{0}^{-\frac{D-2k}{s}}%
\delta^{\left(  k\right)  }\tilde{R}^{\left(  k\right)  }  +a_{k}\xi^{\frac{D-2k}{s}}\tilde{\phi}^{m_{k}}\delta^{\left(  k\right)
}\tilde{S}^{\left(  k\right)  }  \right)  \,.
\end{align}

Therefore, from solutions of the field equations coming from
(\ref{shortaction}) defined by the pair $\left(  g_{\mu\nu},\phi\right)  $,
one can construct solutions of the dual theory with action
(\ref{shortactiondual}), which are defined by%
\begin{equation}
\tilde{g}_{\mu\nu}   =\left(  \frac{\phi _0}{\phi }\right)^{\frac{2}{s}} g_{\mu\nu
} , \ \ \ \ \ \ \ \ \tilde{\phi}  =\xi \frac{\phi_{0}}{\phi  } \,.
\end{equation}

Observe that, as mentioned at the end of section 2, the duality transformation that maps (\ref{shortaction}) into
(\ref{shortactiondual}) interchanges the matter and gravity parts. Below, we
will see how this duality works in some examples, allowing one to find new
solutions within the same family of theories, but in general modifying the
value of the coupling constants. To make the discussion as concrete as
possible, let us consider the familiar action
\begin{equation}
I\left[  g,\phi \right]  =\int d^{4}x\sqrt{-g}\left(  \frac{1}{16\pi
G}R-\frac{\Lambda}{8\pi G}-\partial_{\mu}\phi \partial^{\mu}\phi-\frac{1}{6}%
R\phi^{2}-\frac{\lambda}{4!}\phi^{4}\right)  \,,\label{action4d}%
\end{equation}
which corresponds to the action (2.8) in dimension $D=4$. That is, we
have chosen the conformal weight of the scalar to be $s=-1$. This allows
allows to have a canonically normalized kinetic term for the scalar and, automatically, the conformal coupling with the curvature in its usual form.

\subsection{Self-dual gravitational wave solutions}

The first type of solution we will consider is the analogue of $pp$-waves in AdS space; the so-called AdS-waves. Actually, the theory defined by Eq. (\ref{action4d}) admits the following AdS-wave solution\footnote{Note
that if $h\left(  u,x,y\right)  =f_{1}\left(  u\right)  \left(  x^{2}%
+y^{2}\right)  +f_{2}\left(  u\right)  x+f_{3}\left(  u\right)  $ then metric
is locally AdS$_4$.}%
\begin{equation}
ds^{2}=\frac{l^{2}}{y^{2}}\left[  -f\left(  u,x,y\right)  du^{2}%
-2dudv+dx^{2}+dy^{2}\right]  \,,
\end{equation}
where the scalar field configuration is given by%
\begin{equation}
\phi=\pm2\sqrt{\frac{6}{\lambda}}\frac{1}{l}\frac{y}{x} \,,\label{field}%
\end{equation}
and the profile function $f$ solves the following master equation%
\begin{equation}
-128Gy^{3}\pi\partial_{x}f+2x^{3}l^{2}\lambda \partial_{y}f+xy(64G\pi
y^{2}-\lambda l^{2}x^{2})\left(  \partial_{xx}^{2}f+\partial_{yy}^{2}f\right)
=0 \, .
\end{equation}
(See \cite{AyonBeato:2006jf} for a detailed discussion on AdS-waves supported by a non-minimally coupled scalar). The equation is not separable in these coordinates; nevertheless,
one can go to polar-like coordinates in which the resulting equation turns out to be separable in products. The solution, in terms of the original coordinates $x$ and $y$, reads%
\begin{align}
f\left(  u,x,y\right)    =G\left(  u\right)  & \left[  c_{1}\left(
x^{2}+y^{2}\right)  ^{\frac{1+\sqrt{1-\eta}}{2}}+c_{2}\left(  x^{2}%
+y^{2}\right)  ^{\frac{1-\sqrt{1-\eta}}{2}}\right]  \times\nonumber\\
&  \left[  c_{3}H\left(  a,b,c,d,-\frac{1}{2},-\frac{1}{2},\frac{x^{2}}%
{x^{2}+y^{2}}\right)  +c_{4}\left(  \frac{x^{2}}{x^{2}+y^{2}}\right)
^{\frac{3}{2}}\right.  \times\nonumber\\
&  \left.  \times H\left(  a,b-\frac{3}{4}a+3+\frac{3}{2}c+\frac{3}%
{2}d,c+\frac{3}{2},d+\frac{3}{2},\frac{5}{2},-\frac{1}{2},\frac{x^{2}}%
{x^{2}+y^{2}}\right)  \right]  \,, \label{solheun}%
\end{align}
where $H$ is the general Heun function with parameters%
\be
a=\frac{k}{k+\lambda l^{2}} \,,\ b=-\frac{\eta k}{4k+4\lambda l^{2}} \,,
\ c=-\frac{1}{2}+\frac{1}{2}\sqrt{1-\eta} \,,\ d=-\frac{1}{2}\frac{1-\eta
+\sqrt{1-\eta}}{\sqrt{1-\eta}} \,.
\ee
In \eqref{solheun}, $G\left(  u\right)  $ is an arbitrary function, $\eta$ appears as the
separation constant, $k\equiv 64\pi G$ and $c_{i}$ are integration constants. For
the special case $\eta=1$, the solution develops a logarithmic branch and the
first parethesis of (\ref{solheun}) has to be replaced by the expression
\begin{equation}
c_{1}\sqrt{x^{2}+y^{2}}+c_{2}\sqrt{x^{2}+y^{2}}\log\left(  \sqrt{x^{2}+y^{2}}\right) 
\,.
\end{equation}

There are two particular cases in which the equation reduces further and
admits to be expressed in terms of simpler special functions. Such
cases correspond to $l^{2}\lambda=\pm64\pi G$. For $l^{2}\lambda=64\pi G$ the
solution can be written in terms of hypergeometric functions $_{2}F_{1}$ as
%\begin{align}
%f\left(  u,x,y\right)   &  =G\left(  u\right)  \left[  c_{1}\left(
%x^{2}+y^{2}\right)  ^{\frac{1+\sqrt{1-\eta}}{2}}+c_{2}\left(  x^{2}%
%+y^{2}\right)  ^{\frac{1-\sqrt{1-\eta}}{2}}\right]  \times\nonumber\\
%&  \left[  c_{3}\ {} _{1}F_{2}\left(  A,B,C,\frac{4x^{2}y^{2}}{\left(  x^{2}+y^{2}\right)
%^{2}}\right)  +c_{4}\left(  \frac{4x^{2}y^{2}}{\left(  x^{2}+y^{2}\right)
%^{2}}\right)  ^{1-C}\right.  \times\\
%&  \left.  \times {} _{1}F_{2}\left(  1+A-C,1+B-C,2-C,\frac{4x^{2}y^{2}}{\left(
%x^{2}+y^{2}\right)  ^{2}}\right)  \right]  \ ,
%\end{align}
\begin{align}
f\left(  u,x,y\right)    =G\left(  u\right)  &\left[  c_{1}\left(
x^{2}+y^{2}\right)  ^{\frac{1+\sqrt{1-\eta}}{2}}+c_{2}\left(  x^{2}%
+y^{2}\right)  ^{\frac{1-\sqrt{1-\eta}}{2}}\right]  \times\nonumber\\
&  \left[  c_{3}\ {} _{2}F_{1}\left(  A,B,C,\frac{4x^{2}y^{2}}{\left(  x^{2}+y^{2}\right)
^{2}}\right)  +c_{4}\left(  \frac{4x^{2}y^{2}}{\left(  x^{2}+y^{2}\right)
^{2}}\right)  ^{1-C}\right.  \times \nonumber\\
&  \left.  \times {} _{2}F_{1}\left(  1+A-C,1+B-C,2-C,\frac{4x^{2}y^{2}}{\left(
x^{2}+y^{2}\right)  ^{2}}\right)  \right]  \,,
\end{align}
with%
\be
A=-\frac{1}{4}\left(  1+\sqrt{1-\eta}\right)  \,, \ B=-\frac{1}{4}\left(
1-\sqrt{1-\eta}\right)  \,, \ C=-\frac{1}{2} \,,
\ee
while for $l^{2}\lambda=-64\pi G$ the equation for the profile integrates in
terms of generalized Legendre functions as%
%\begin{align}
%f\left(  u,x,y\right)   &  =G\left(  u\right)  \left[  c_{1}\left(
%x^{2}+y^{2}\right)  ^{\frac{1+\sqrt{1-\eta}}{2}}+c_{2}\left(  x^{2}%
%+y^{2}\right)  ^{\frac{1-\sqrt{1-\eta}}{2}}\right]  \times\nonumber\\
%&  \left(  \frac{xy}{x^{2}+y^{2}}\right)  ^{3/2}\left[  c_{3}P_{\nu}^{\mu
%}\left(  \frac{x^{2}-y^{2}}{x^{2}+y^{2}}\right)  +c_{4}P_{\nu}^{-\mu}\left(
%\frac{x^{2}-y^{2}}{x^{2}+y^{2}}\right)  \right]  \ ,
%\end{align}
\begin{align}
f\left(  u,x,y\right)   =G\left(  u\right)  &\left[  c_{1}\left(
x^{2}+y^{2}\right)  ^{\frac{1+\sqrt{1-\eta}}{2}}+c_{2}\left(  x^{2}%
+y^{2}\right)  ^{\frac{1-\sqrt{1-\eta}}{2}}\right]  \times\nonumber\\
&  \left(  \frac{xy}{x^{2}+y^{2}}\right)  ^{3/2}\left[  c_{3}P_{\nu}^{\mu
}\left(  \frac{x^{2}-y^{2}}{x^{2}+y^{2}}\right)  +c_{4}P_{\nu}^{-\mu}\left(
\frac{x^{2}-y^{2}}{x^{2}+y^{2}}\right)  \right]  \,,
\end{align}
where $P_{\nu}^{\mu}$ are the associated Legendre functions of first kind and%
\begin{equation}
\nu=\frac{1}{2}\left(  \sqrt{4-\eta}-1\right)  \ , \ \ \ \ \ \mu=\frac{3}{2} \,.
\end{equation}
In both cases, the solution develops a logarithmic branch for $\eta=1$.

Now, we can consider the configurations dual to these solutions. For $l^{2}%
\lambda=64\pi G$ the metric configuration obtained after dualization $g\to \tilde{g} $ is
locally equivalent to the original one. Furthermore, at this point of the parameter space, the value of the action remains
invariant under such dualization. In this sense, this configuration can be referred to as
{\it self-dual} with no ambiguity. For $l^{2}\lambda=-64\pi G$, on the other hand, the configuration
obtained after $g\to \tilde{g}$ is also locally equivalent to the original one; however, in this case the action is mapped to minus itself, and thus this case can be regarded as {\it anti-self-dual}.

\subsection{Self-dual black hole solutions}

Now, let us go back to black holes and again consider the 4-dimensional theory (\ref{theoryaction}). As shown in \cite{Martinez2}, provided certain relation between the coupling constants, the following metric and scalar field represent a solution of the theory (\ref{action4d})%
%\begin{align}\label{MTZ}
%ds^{2}  & =-\left(  -\frac{\Lambda}{3}r^{2}+K\left(  1+\frac{\mu}{r}\right)
%^{2}\right)  dt^{2}+\frac{dr^{2}}{\left(  -\frac{\Lambda}{3}r^{2}+K\left(
%1+\frac{\mu}{r}\right)  ^{2}\right)  }+r^{2}d\Sigma_{K}^{2} \,, \nonumber\\
%\phi & =\sqrt{\frac{3}{8\pi G}}\frac{\mu}{r+\mu}\ ,
%\end{align}
\begin{equation}\label{MTZ}
ds^{2}  =-\left(  -\frac{\Lambda}{3}r^{2}+\gamma\left(  1+\frac{\mu}{r}\right)
^{2}\right)  dt^{2} + \left(-\frac{\Lambda}{3}r^{2}+\gamma\left(  1+\frac{\mu}{r}\right)  ^{2}\right)^{-1} dr^2 + r^{2}d\Sigma_{\gamma}^{2} \,,
\end{equation}
and
\begin{equation}
\phi  =\sqrt{\frac{3}{8\pi G}}\frac{\mu}{r+\mu} \,,
\end{equation}
where $d\Sigma_{\gamma}$ is the line element of a two-dimensional space of constant
curvature $\gamma$ normalized to $\pm1,0$, and $\mu$ is an integration constant. For this to be a solution, the
coupling of the conformal potential has to be related to the bare cosmological
constant and Newton's constant by the equation
\begin{equation}
\lambda=-\frac{64\pi}{3}G\Lambda \,.\label{constMTZ}%
\end{equation}

This equation relates the matter coupling $b_{0}\sim\lambda$ with the
gravitational coupling $a_{0}\sim\Lambda$. By performing the transformation $g\to \tilde{g}$, the dual solution is observed to be
%\begin{align*}
%d\tilde{s}^{2}  & =\phi_{0}^{-2}\left(  \sqrt{\frac{3}{8\pi G}}\frac{\mu
%}{r+\mu}\right)  ^{2}\left(  -\left(  -\frac{\Lambda}{3}r^{2}+K\left(
%1+\frac{\mu}{r}\right)  ^{2}\right)  dt^{2}+\frac{dr^{2}}{\left(
%-\frac{\Lambda}{3}r^{2}+K\left(  1+\frac{\mu}{r}\right)  ^{2}\right)  }%
%+r^{2}d\Sigma_{K}^{2}\right)  \text{ ,}\\
%\tilde{\phi} & =\phi_{0}\xi\sqrt{\frac{8\pi G}{3}}\frac{r+\mu}{\mu}\ ,
%\end{align*}
\begin{alignat}{3}
d\tilde{s}^{2}  & =   {\frac{3}{8\pi G}} \left(  \frac{\mu / \phi_{0}
}{r+\mu}\right)  ^{2} \Bigg[  -\Big( && -\frac{\Lambda}{3}r^{2} + \gamma\left(
1+\frac{\mu}{r}\right)  ^{2}\Big)  dt^{2} + \nonumber\\
& && \left(-\frac{\Lambda}{3}r^{2}+\gamma\left(  1+\frac{\mu}{r}\right)  ^{2}\right)^{-1} dr^2
+r^{2}d\Sigma_{\gamma}^{2} \Bigg]  \,, 
\end{alignat}
with
\begin{equation}
\tilde{\phi}  =\phi_{0}\xi\sqrt{\frac{8\pi G}{3}}\frac{r+\mu}{\mu} \,, 
\end{equation}
which, after choosing the free parameters $\phi_{0}$ and $\xi$ as%
\begin{equation}
\phi_{0}=\xi=\sqrt{\frac{3}{8\pi \, G}}\ ,
\end{equation}
rescaling the time, and changing the radial coordinate as%
\begin{equation}
\rho= \frac{\mu r}{r+\mu} \,,
\end{equation}
reduces to%
%\begin{align}
%d\tilde{s}^{2}  & =-\left(  -\frac{\Lambda}{3}\rho^{2}+K\left(  1+\frac
%{\tilde{\mu}}{\rho}\right)  ^{2}\right)  dt^{2}
%+ \frac{d\rho^{2}}{\left(
%-\frac{\Lambda}{3}\rho^{2}+K\left(  1+\frac{\tilde{\mu}}{\rho}\right)
%^{2}\right)  }+\rho^{2}d\Sigma_{K}^{2}\ ,\\
%\tilde{\phi} & =\sqrt{\frac{3}{8\pi G}}\frac{\tilde{\mu}}{\rho+\tilde{\mu}}\ .
%\end{align}
\begin{equation}
d\tilde{s}^{2}   =-\left(  -\frac{\Lambda}{3}\rho^{2}+\gamma\left(  1+\frac
{\tilde{\mu}}{\rho}\right)  ^{2}\right)  dt^{2}
+ \left( -\frac{\Lambda}{3}\rho^{2}+\gamma\left(  1+\frac{\tilde{\mu}}{\rho}\right)
^{2}\right) d\rho^{2} + \rho^{2}d\Sigma_{\gamma}^{2} \,, 
\end{equation}
with 
\begin{equation}
\tilde{\phi}  =\sqrt{\frac{3}{8\pi G}}\frac{\tilde{\mu}}{\rho+\tilde{\mu}} \,.
\end{equation}

Therefore, we observe that after dualization of the black hole solution (\ref{MTZ}) one obtains, locally, the
original configuration; what can be seen explicitly by fixing the arbitrary
parameters $\phi_{0}$ and $\xi$ and redefining $\mu=-\tilde{\mu}$. Since the solution gets mapped to a solution that is diffeomorphic to itself, it can be
regarded as a self-dual solution as well. Under these conditions and using
(\ref{constMTZ}), one can see that the action maps to itself as well, as it occurs for one of the two AdS-wave solutions considered above. 

It is worth emphasizing that in $D>4$ dimensions, the black hole solutions are not necessarily self-dual in the sense discussed here. Let us comment on the dual geometry corresponding to the higher-dimensional black hole solution; that is, the geometry described by the metric $\tilde{g}$ when $g$ is given by (\ref{solutionhigher}). Since, for $s=-1$ and for arbitrary $D>4$ the scalar field configuration is given by $\phi = N/r$, the rescaled metric turns out to be
\be
d\tilde{s}^2=-\frac{N^2F(r)}{r^2} dt^2 + \frac{N^2}{F(r)r^2} dr^2 + N^2 d\Sigma_{D-2}^2 \,.\label{dualingo}
\ee

We observe from this that the dual geometry correspond to the product of a 2-dimensional space ${\mathcal M}_2$ and a $(D-2)$-sphere of radius $N$. Then, since $\phi(r)$ is regular for all $r\neq 0$, the null surfaces are still located at $r=r_+$. The asymptotic (large $r$) behaviour of the 2-dimensional space ${\mathcal M}_2$ depends on the value of $\Lambda $ in the action. While for $\Lambda <0 $ the space (\ref{dualingo}) tends asymptotically to $\mathbb{R}^{1,1}\times S^{D-2}$, for $\Lambda =0 $ it turns out to be asymptotically AdS$_2 \times S^{D-2}$, where both the radius of AdS and of the sphere are given by $N$.

\begin{acknowledgments}

The authors thank Sourya Ray for useful discussions and previous collaboration in this subject. We also want to thank Alberto G\"uijosa for useful comments on the manuscript. This work has been supported by FONDECYT
Grants 1141073 and 1150246. The support of Fundaci\'on Bunge y Born, Universidad de Buenos Aires and CONICET is acknowledged. This work has also been partially funded by FNRS-Belgium (convention FRFC PDR T.1025.14 and convention IISN 4.4503.15), by CONICET of Argentina through grant PIP0595/13, by the Communaut\'{e} Fran\c{c}aise de Belgique through the ARC program and by a donation from the Solvay family. MC is partially supported by Mexico's National Council of Science
and Technology (CONACyT) grant 238734 and DGAPA-UNAM grant IN113115.

\end{acknowledgments}

\newpage

%\appendix
%
%\section{Awful formalae}
%\label{app}

%\bibliography{HairyBib}{}
%\bibliographystyle{JHEP}

\end{document}